\documentclass{aastex631}
\usepackage{xcolor}

\newcommand\kms{\,${\rm km\>s}^{-1}$\>}
\newcommand\kmsns{\,${\rm km\>s}^{-1}$}
\def\up#1{\leavemode \raise.16ex\hbox{#1}}

\begin{document}

\title{Detection of Rydberg lines from the atmosphere of Betelgeuse}

\author[0000-0002-2490-1079]{W. R. F. Dent}
\affiliation{ALMA Santiago Central Offices, Alonso de Cordova 3107, Vitacura, Casilla 763 0355,  Santiago, Chile}

\author[0000-0002-7042-4541]{G. Harper}
\affiliation{Center for Astrophysics and Space Astronomy, University of Colorado Boulder, USA}

\author{A. M. S. Richards}
\affiliation{Jodrell Bank Centre for Astrophysics, University of Manchester, Manchester, UK}

\author[0000-0003-0626-1749]{P. Kervella}
\affiliation{LESIA, Observatoire de Paris, Université PSL, CNRS, Sorbonne Université, Université Paris-Cité, 5 Place Jules Janssen, 92195 Meudon, France}

\author{L. D. Matthews}
\affiliation{MIT Haystack Observatory, 99 Millstone Road, Westford, MA 01886 USA}

\begin{abstract}
Emission lines from Rydberg transitions are detected for the first time from a region close to the surface of Betelgeuse. The H30$\alpha$ line is observed at 231.905~GHz, with a FWHM $\sim42$\,\kms and extended wings. A second line at 232.025~GHz (FWHM $\sim21$\kms),
is modeled as a combination of Rydberg transitions of abundant low First Ionization Potential metals. 
Both H30$\alpha$ and the Rydberg combined line X30$\alpha$ are fitted by Voigt profiles, and collisional broadening with electrons may be partly responsible for the Lorentzian contribution, indicating electron densities of a few 10$^8$ cm$^{-3}$.  X30$\alpha$ is located in a relatively smooth ring at a projected radius of 0.9$\times$ the optical photospheric radius $R_{\star}$, 
whereas H30$\alpha$ is more clumpy, reaching a peak at $\sim1.4 R_{\star}$. 
We use a semi-empirical thermodynamic atmospheric model of Betelgeuse to compute the 232 GHz (1.29\,mm) continuum and line profiles making simple assumptions. Photoionized abundant metals dominate the electron density and the predicted surface of continuum optical depth unity at 232 GHz occurs at $\sim1.3R_\star$, in good agreement with observations. Assuming a Saha-Boltzmann distribution for the level populations of Mg, Si, and Fe, the model predicts that the X30$\alpha$ emission arises in a region of radially-increasing temperature and turbulence. Inclusion of ionized C and non-LTE effects could modify
the integrated fluxes and location of emission.  These simulations confirm the identity of the Rydberg transition lines observed towards Betelgeuse, and reveal that such diagnostics can improve future atmospheric models. 

\end{abstract}



\section{Introduction} \label{sec:intro}

Red supergiants (RSGs) are an important late stage in the evolution of massive stars. Their high mass loss rates are one of the major sources of enrichment of the ISM, yet the details of the mass loss mechanism are not fully understood. Interaction of granulation,  convective cells and pulsations with magnetic fields, in addition to scattering by lines or nascent dust followed by radiation pressure are mechanisms proposed to initiate mass loss and drive the wind \citep{Airapetian2000, Arroyo-Torres2015, 2009A&A...498..127V, 2021ARA&A..59..337D}. 
Episodic and inhomogeneous ejection  may also play a significant role (\citealt{Montarges2021}, \citealt{Humphreys2022}). High-resolution infrared and mm observations of the closest RSGs show molecular gas - mostly observed through CO, SiO and H$_2$O - lying in a highly-clumped shell (or MOLsphere) at radii of $\sim1.5R_{\star}$ \citep{Perrin2007}. Radio through mm continuum spectral energy distributions show optically thick free-free emission arising from a hot chromosphere with a small filling factor embedded in more pervasive lukewarm plasma heated to 2000--3600K, and this extends from $1.3R_{\star}$ out to tens of R$_*$ at the longer (cm) wavelengths (\citealt{Lim1998}, \citealt{O'Gorman2020}). How the clumpy MOLsphere and the chromosphere are related and involved in the mass loss from these stars remains unclear.

At a distance of 222~pc \citep{Harper2017b}, Betelgeuse is the second-closest RSG after Antares, and consequently has been extensively studied as an archetype (e.g.~\citealt{Wheeler2023}). Its K-band (2.2$\micron$) spectro-interferometric angular diameter of 42.5\,mas \citep{2011A&A...529A.163O} makes it resolvable with current IR and radio instruments. 
\cite{Harper2017a} found Betelgeuse's center-of-mass radial velocity to be
$V_{\mathrm{helio}}$=($20.9\pm  0.3$)~\kms based on a mean of four different diagnostics, and \cite{Kervella2018} measured $V_{\mathrm{helio}}$=($20.4\pm  0.1$)~\kms from ALMA SiO emission. We adopt $V_{\star} = V_{\mathrm{helio}}$=20.9~\kms (corresponding to $V_{\mathrm{lsr}}$ of 4.9~\kms), and $R_*$ = 1014 R$_{\odot}$. 
The structure is not symmetrical - continuum hotspots in radio through mm and in infrared continuum are seen on scales of hundreds down to $\sim15$mas (\citealt{Richards2013}, \citealt{O'Gorman2020}, \citealt{Haubois2009}), and macroscopic gas motions have been imaged in the clumpy gas on scales down to tens of mas - within 0.5 $R_{\star}$ of the photosphere (e.g., \citealt{Montarges2014}, \citealt{Kervella2018}). In 2020, Betelgeuse underwent a historically-significant 1.7 magnitude drop in V-band brightness, ascribed either to ejection of an obscuring dust clump along the line of sight, or the formation of cool regions on the photosphere correlated with the pulsating atmosphere, or both (\citealt{Montarges2021}, \citealt{Harper2020}). This has stirred up considerable new interest in this star.

Rydberg transition lines (RTLs) of Hydrogen result from changes in the principal quantum number {\it n}. As well as H and He, RTLs of heavier elements such as C, Si etc have also been observed, but because such atoms with electrons in high energy levels are hydrogenic, their lines appear close in frequency. In environments such as HII regions, planetary nebulae, photodissociation regions and young stellar objects, Rydberg transitions result from radiative recombination of the ions, and \cite{Gordon2002} provide an extensive review of the resulting radio recombination lines.  \citet{Olofsson2021} recognised that the emission around 232.02~GHz in the binary AGB star HD101584 is likely due to a superposition of 30$\alpha$ RTLs from elements heavier than C, for which they coined the term X30$\alpha$. Their model indicated that Mg30$\alpha$ may dominate and, moreover, the environment is such that these lines may not be solely due to radiative recombination (hence the more general term RTL). Infrared RTLs from Mg~I as high as $n=8$ have also been observed and modelled in the Sun \citep{Chang1991,Carlsson1992} and 
cool evolved stars \citep{Uitenbroek1996,Sundqvist2008}.

In this work, we present the first detection of RTLs in the atmosphere of Betelgeuse, at $\sim$232~GHz, using ALMA. These observations were obtained as part of a multi-band high-resolution program to study this star after the Great Dimming event,  providing images of the continuum and a number of molecular lines. Full results will be presented in a separate paper. 

\section{Observations} \label{sec:observations}

Betelgeuse was observed 
using ALMA in its most extended configuration giving a maximum baseline of 15~km, in
two executions on Aug 3 and Aug 27 2023.  
We used a total bandwidth of 7.5~GHz in four spectral windows centred at 214.769, 217.073, 220.266 and 231.966~GHz (1.40--1.29 mm) with a spectral channel spacing of 0.976~MHz and an effective spectral resolution of 2.6~\kms at 232~GHz after Hanning smoothing. Standard phase referencing was used, cycling to the phase calibrator J0552+0313 (at 4.2$^{\circ}$ separation) every 80--90 seconds. The bandpass was calibrated using J0510+1800. 
The standard calibrated measurement sets (visibility data) from the ALMA {\it CASA} pipeline were concatenated, and the line-free channels used to made a continuum image to provide a model for phase self-calibration in {\it CASA} \citep{2022PASP..134k4501C}. 
The final continuum image was made solving for the stellar spectral index, using a restoring beam of ($27\times19$)\,mas FWHM at PA~$51^{\circ}$, achieving an off-source continuum noise level of  0.024~mJy beam$^{-1}$.
The calibration was applied to all spectral channels and we made spectral image cubes before and after continuum subtraction. 
The restoring beam at around 232~GHz is ($28.8\times20.6$)\,mas FWHM at PA~$50^{\circ}$ and the noise level per spectral channel is $\sigma_{\mathrm {rms}} =0.9$ mJy beam$^{-1}$. The maximum recoverable scale is 0\farcs35. The flux scale is accurate to $\sim7\%$ and the astrometric accuracy is $\sim$3~mas.

\section{Results} \label{sec:results}

The ALMA resolution around 232 GHz provides $\sim$two resolution elements across the photosphere, and a 2D Gaussian component fitted to the 214--233 GHz continuum  has a radius at FWHM (after deconvolution from the beam) of $R_{\mathrm {232GHz}}$ = 30\,mas.  
We measure the stellar centre as ICRS 05:55:10.3435 +07:24:25.640 with $\sim$3~mas stochastic uncertainty.
In the image plane (convolved with the beam) we found that an aperture of radius 65~mas ($R_{65}$) enclosed all the emission, both line and continuum.
The disc-average spectral index between 214--233 GHz ($\alpha$) is approximately 1.7 and the continuum flux density around 232 GHz is 297$\pm$20 mJy in $R_{65}$. 

A number of lines of SiO and CO are covered by this project - these will be discussed in a later paper - but here we focus on two bright emission lines found 
around 232~GHz, which we identify as 30$\alpha$ Rydberg transitions. 

\begin{figure}
\begin{center}
	\includegraphics[width=8.5cm]{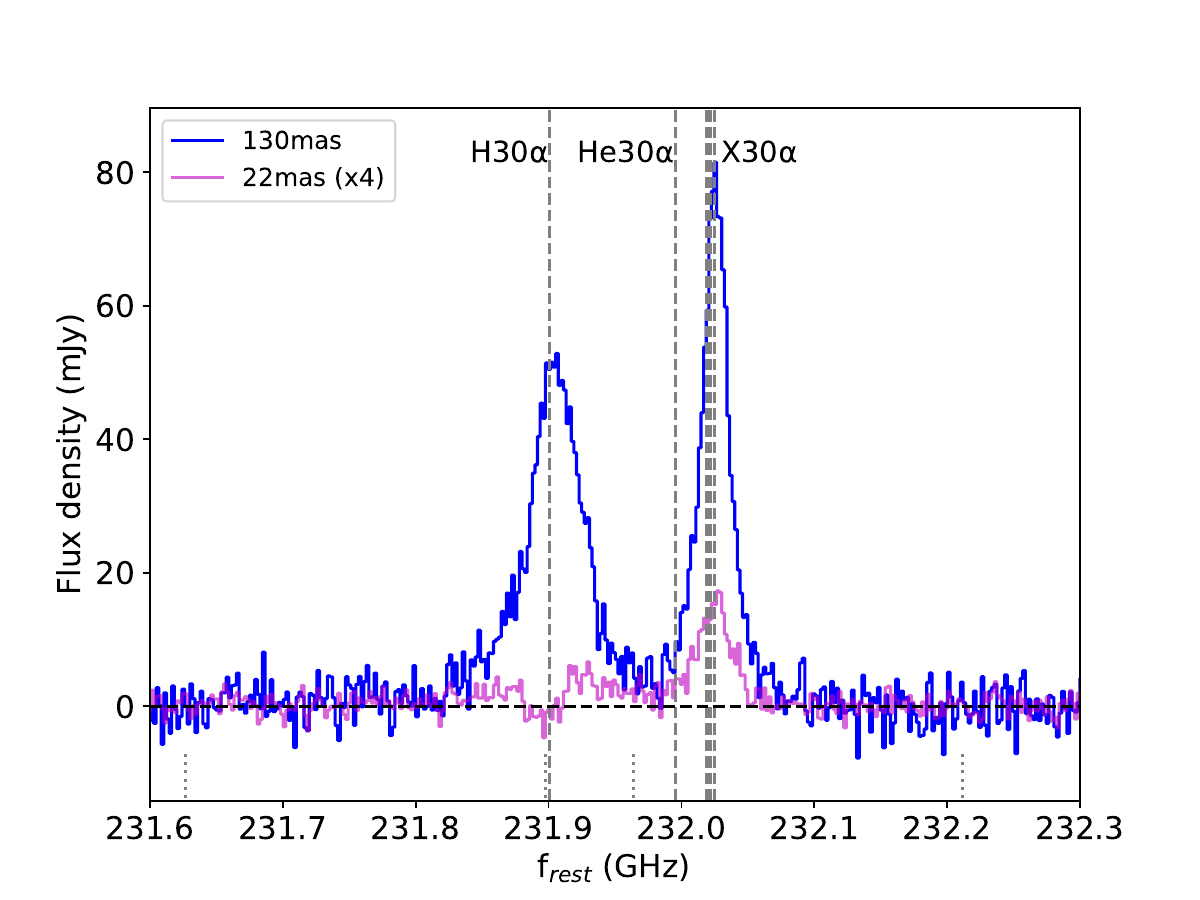}
 	\includegraphics[width=8.5cm]{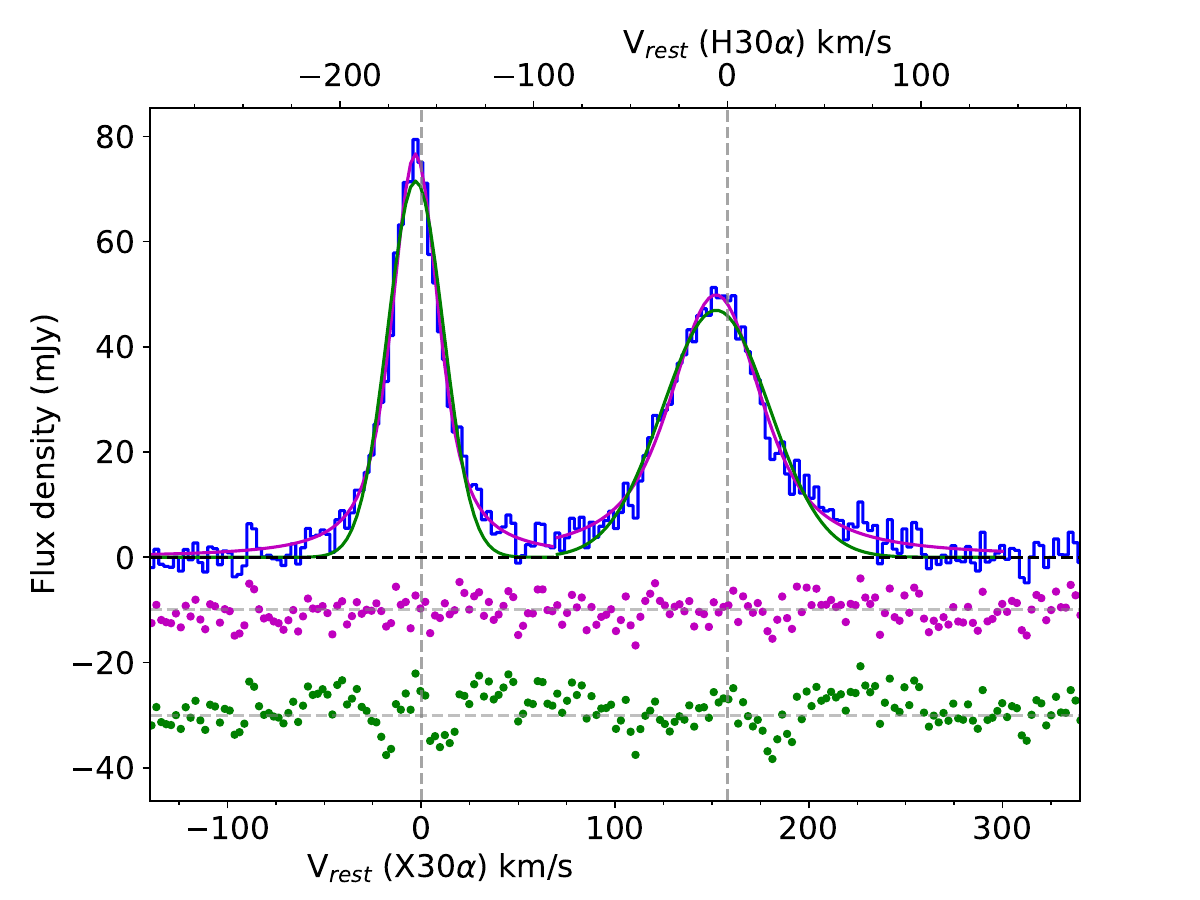}
    \caption{Spectrum of 30$\alpha$ Rydberg lines from Betelgeuse. Left panel shows emission after continuum subtraction, integrated over a 130\,mas aperture (blue) and from only the central 22\,mas (scaled by $\times$4, magenta).  The frequencies of H30$\alpha$, He30$\alpha$ and the group of O, Mg, S and Fe (in order of increasing Rydberg frequency, and marked X30$\alpha$) are indicated by the vertical dashed lines. Also shown by short dotted lines are the line frequencies of TiO$_2$. Right panel shows zoom in after binning by two channels, on velocity scales based on the calculated H30$\alpha$ and X30$\alpha$ rest frequencies in Table~\ref{tab:line_chars}.  Voigt profiles (magenta lines) provide good fits to the line core and wings while Gaussian profiles (green) under-predict the line wings. Residuals for the two types of profiles are shown offset below. Details of the fits are given in Table~\ref{tab:line_chars}. Axes in both cases are adjusted to the rest frame with respect to $V_{\star}$.
    }
    \label{fig:recomb_lines}
    \end{center}
\end{figure}

Fig.~\ref{fig:recomb_lines} (left panel) shows the continuum-subtracted integrated spectrum taken within $R_{65}$. 
Frequencies are shown in the stellar rest frame, and the rest frequencies of some 30$\alpha$ RTLs are indicated.  We also show TiO$_2$ frequencies covered by this spectrum, as this molecule has been detected in some other RSGs (e.g.~\citealt{DeBeck2015}). However, based on the relative CDMS/JPL intensities listed in {\it Splatalogue}\footnote{https://splatalogue.online/}, the most prominent TiO$_2$ lines should be at 231.626 and 231.963~GHz, and we conclude that there is no evidence of significant contribution from this species.  No other obvious candidates for these lines could be found. 
The characteristics of the observed lines and comparison with expected line frequencies are given in Table~\ref{tab:line_chars}, including an
upper limit for the He30$\alpha$ line. 
The brightest peak is centred close to the cluster of 30$\alpha$ RTLs from high abundance elements including Mg, Si, Fe, and perhaps O and S. This is similar to the cluster of lines noted by \citet{Olofsson2021} in HD101584.
The left panel in Fig.~\ref{fig:recomb_lines} illustrates the frequencies of these elements based on the Rydberg formula \citep{Towle1996}. The center frequency of this cluster is affected by the weighting by the fine-structure lines as well as the relative contributions from the different elements, and is discussed further in section \ref{sec:model}.
Although the clustering of lines from the heavier elements make it difficult to 
 separate the dominant species in the X30$\alpha$ group, the emission peaks of both the H30$\alpha$ and X30$\alpha$ lines appear to be blue-shifted relative to $V_{\star}$, by $-5.9$ and $-3.0$~\kms respectively; possible explanations for this will also be discussed in section \ref{sec:model}. Also shown in Fig.~\ref{fig:recomb_lines} (left panel) is the integrated spectrum of only the central 11\,mas radius, $R_{11}$, covering approximately the central beam on the photosphere. Here the H30$\alpha$ line shows a clear dip slightly below the continuum level, indicating absorption centred at $\sim$231.898~GHz. This is $\sim$10\kms red-shifted with respect to the overall emission, with the absorption profile appearing narrower than the main emission component. These observations suggest variations in the gas radial motion and line broadening along the line-of-sight.

\begin{table}
	\centering
	\caption{Line characteristics. }
	\label{tab:line_chars}
	\begin{tabular}{lcccccccc}
		\hline
Line & Calculated freq. & Measured freq.  & Freq. err.  & Width (FWHM) & FWZM & Peak line flux& Total intensity  & Radius \\
 & (GHz) & (GHz)$^3$ & (GHz)$^4$ &  (\kms)$^5$ & (\kms)$^6$ & (mJy)$^7$ & (Wm$^{-2}$)$^7$ & (mas)$^8$ \\
 \hline
H30$\alpha$ & 231.9009$^1$ & 231.9055 & 0.0004 & 42.6 (31.0,19.4) & 180 & 50.6 & $4.5\times10^{-19}$ & 28\\
X30$\alpha$ & 232.0232$^2$ & 232.0255 & 0.0002 & 21.5 (15.3,10.2) & 120 & 80.8 & 5.7$\times10^{-19}$ & 18\\
He30$\alpha$ & 231.9954$^1$  & - & - & - & - & $<$5 & - & -\\
\hline

\end{tabular}

(1) Rest frequency from Rydberg formula \citep{Towle1996};
(2) Combined centre rest frequency (see section \ref{sec:model});
(3) Fitted centre frequency in target frame adjusted to  $V_{\star}=V_{_\mathrm{helio}}$ of 20.9~\kms; 
(4) Formal fitting error;
(5) Combined FWHM, fitted using a Voigt profile (FWHM of individual Gaussian and Lorentzian components are given in brackets);
(6) Full width at approximate noise level, estimated from spectra;
(7) In aperture of 130\,mas diameter;
(8) Projected radius of peak emission, azimuthally-averaged.

\end{table}

The line width of H30$\alpha$ is approximately two times broader than X30$\alpha$. Although there is some contribution to the X30$\alpha$ width from the potential multiple atomic components, Fig.~\ref{fig:recomb_lines} shows that this is relatively small compared with the observed linewidth. 
Both H30$\alpha$ and X30$\alpha$ have broad wings and were better fitted by Voigt rather than Gaussian profiles. This is illustrated in the right panel of Fig.~\ref{fig:recomb_lines}, and the fitted widths (FWHM) of the Gaussian component in both cases was $\sim$3/2 of the Lorentzian (Table~\ref{tab:line_chars}). By comparison, the absorption and emission profiles of SiO have linewidths of typically $\sim$20\kms FWHM \citep{Kervella2018} and optical lines can range from 15\kms, up to $\sim$40\kms in the upper photosphere \citep{Jadlovsky2023}.

\begin{figure}

	\includegraphics[width=7.2cm,angle=-90]{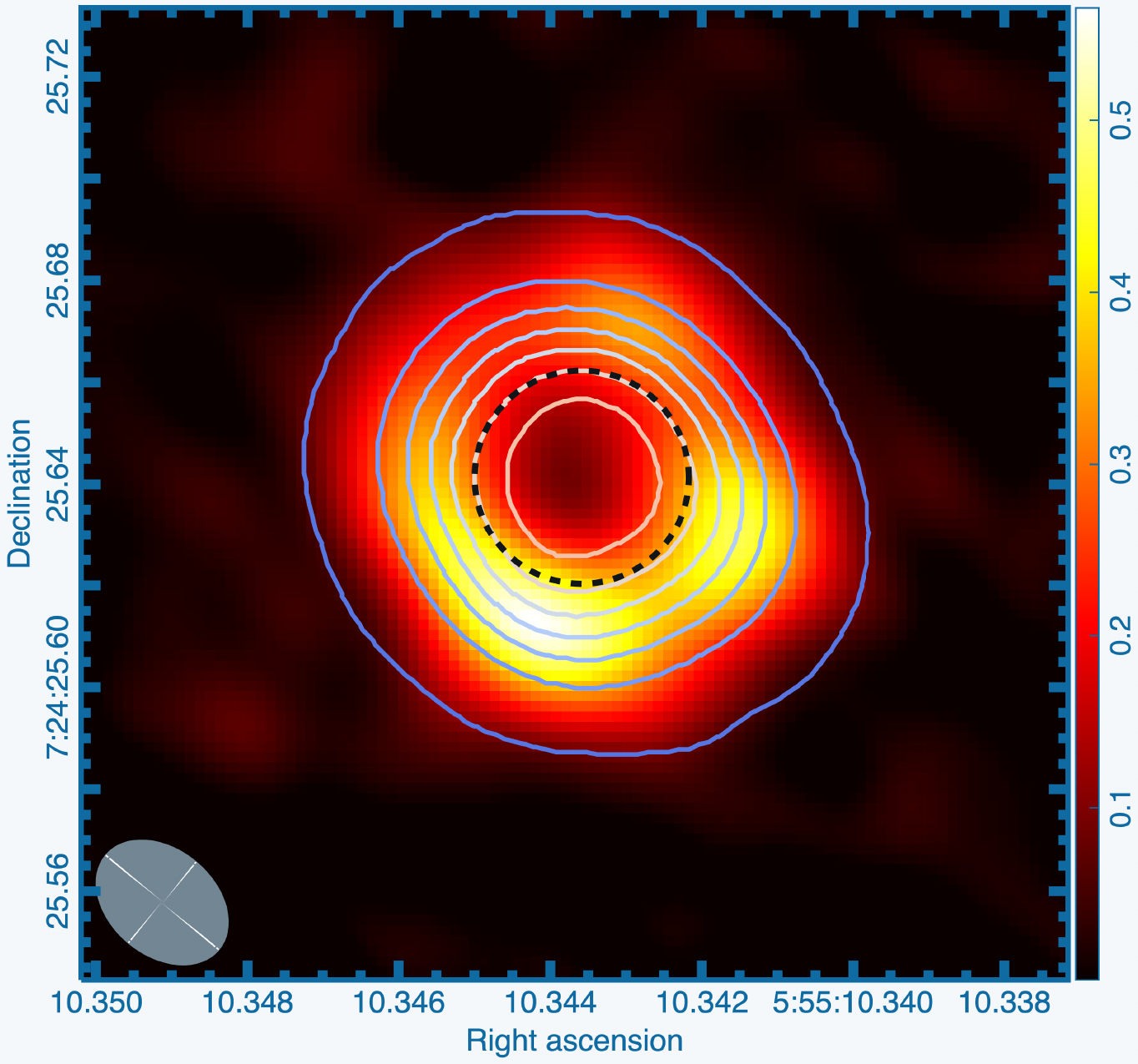}
 	\includegraphics[width=7.2cm,angle=-90]{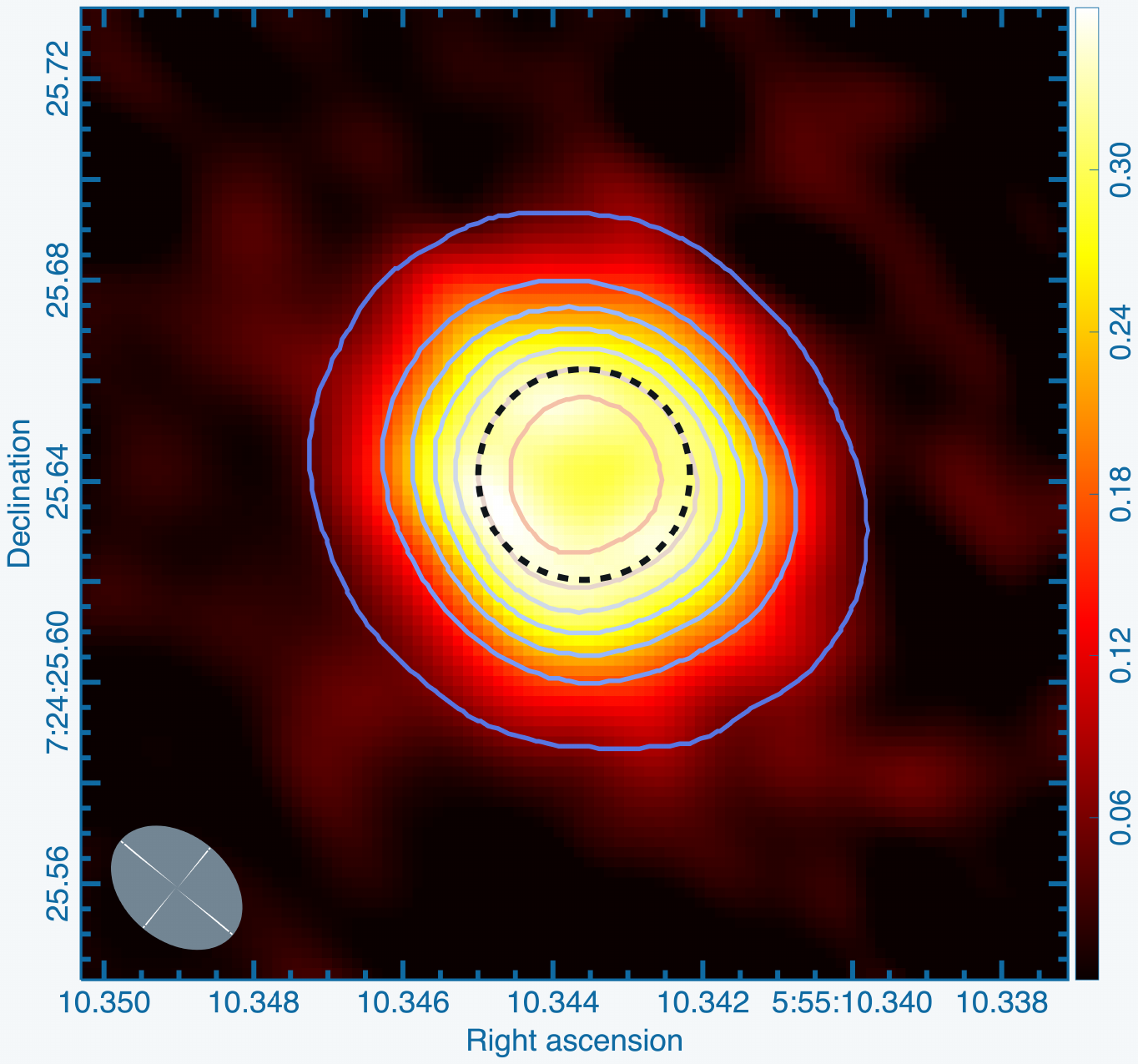}
    \caption{Maps of the continuum-subtracted integrated emission of H30$\alpha$ (231.82 -- 231.97~GHz, left) and X30$\alpha$ (231.97 -- 232.08~GHz, right) in the vicinity of Betelgeuse. The size of the IR photosphere is shown by the dashed circle, and the 214--233 GHz continuum by the contours. Contours start at 1\,mJy beam$^{-1}$, with an interval of 8\,mJy beam$^{-1}$, and the colour scales of line intensity are in Jy beam$^{-1}$ \kmsns. The beam size is shown lower left. }
    \label{fig:recomb_map}

\end{figure}

\begin{figure}
\begin{center}
	\includegraphics[width=8cm]{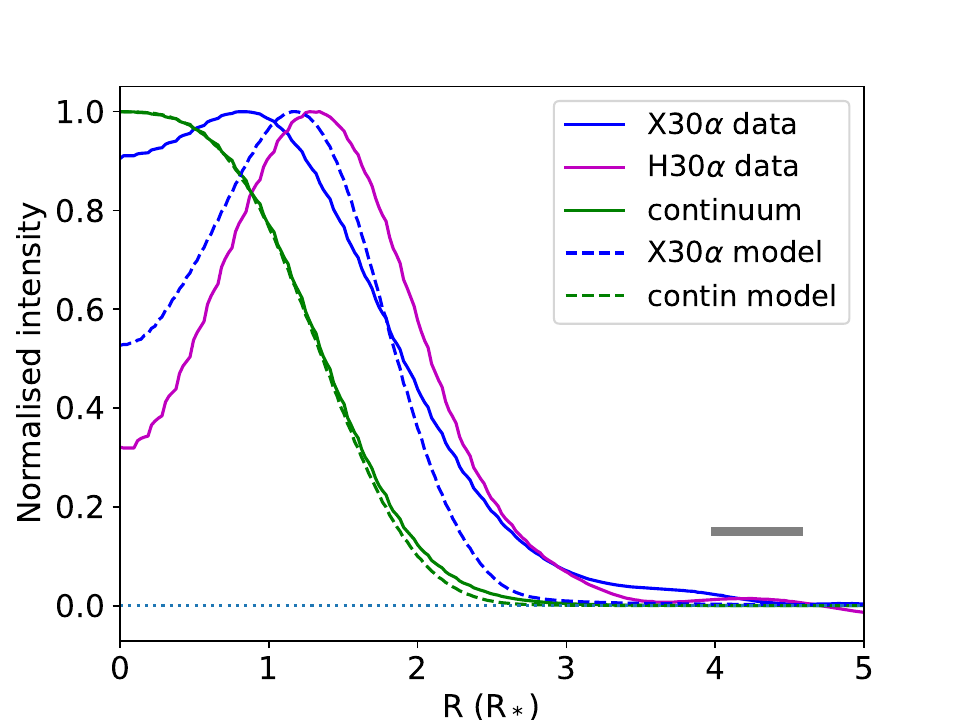}
	\includegraphics[width=9cm]{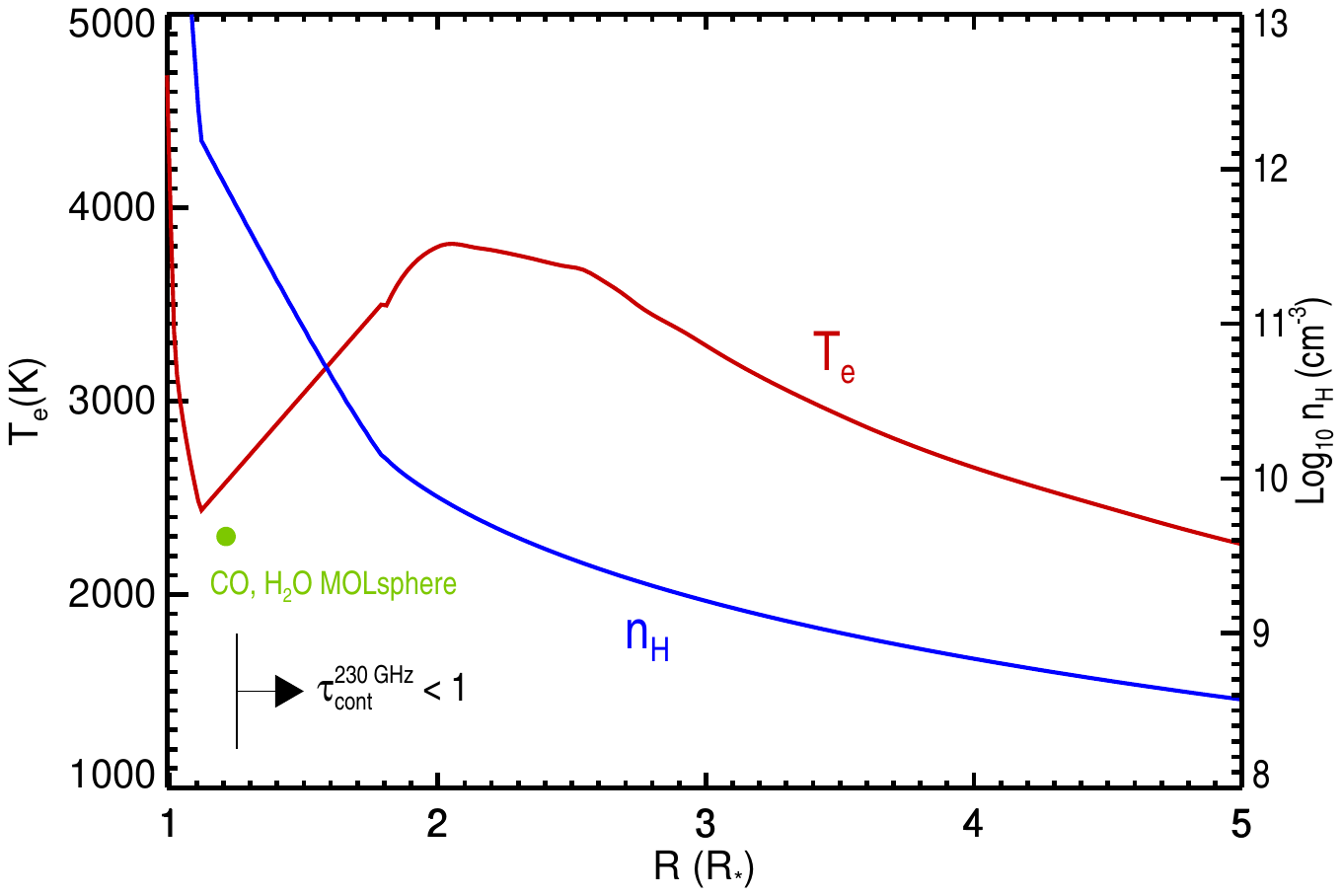}
    \caption{Azimuthally-averaged projected radial distributions. Left panel shows the RTLs (blue and magenta) and the 1.3\,mm continuum (green). Solid lines are the observations and dashed lines represent the model convolved with the azimuthally-averaged ALMA beam (size shown lower right). Right panel illustrates the adopted semi-empirical model structure, including the radial profile of the gas (electron) temperature T$_e$ and neutral H density n$_H$ above the photosphere. Also shown are the inner edge of the MOLsphere and the $\tau = 1$ continuum radius at the observing frequency. The electron density (not shown) is dominated by the photoionization of abundant low
    FIP elements, Mg, Si, and Fe, with contributions from C and S.}
    \label{fig:radial_structure}
    \end{center}
\end{figure}

Fig.~\ref{fig:recomb_map} compares the 
angular distribution of the two
RTLs and 1.3\,mm stellar continuum emission across Betelgeuse, and the left panel of Fig.~\ref{fig:radial_structure} shows the projected radial distributions. The RTLs are generally more extended than the stellar continuum, and lies in ring-like structures of radii 1.4 and 0.9~R$_{\star}$, resembling the molecular shell observed by \citet{Kervella2018} in SiO and CO lines. The molecular shell has a somewhat larger radius ($\sim$~2 R$_*$), although this may partly be affected by blending with the line absorption against the stellar photosphere. 
One characteristic of both RTL images is the central dip towards the photosphere. This is particularly prominent in H30$\alpha$, but in this case it is partly generated by the narrow dip in the central line profile - which is thought to be due to absorption against the bright stellar continuum (left panel in Fig.~\ref{fig:recomb_lines}). 
Note that these are continuum-subtracted line maps, and the lines are mostly seen in emission against a bright optically-thick continuum background. The temperature in the model increases with radius in the emitting region (Fig.~\ref{fig:radial_structure}), which in this case results in the lines being in emission.

X30$\alpha$ emission has more symmetric, less clumpy structure than either  H30$\alpha$ or published SiO maps: azimuthal variations of $\sim10\%$ are found in the X30$\alpha$ map (Fig.~\ref{fig:recomb_map}), which compares with the contrast of 2 or more in the H30$\alpha$ clumps and in other lines \citep{Kervella2018}. This - and the different radial profiles in Fig.~\ref{fig:radial_structure} - indicate that the X30$\alpha$ emission is not arising from the same region as the H30$\alpha$. Closer examination of both datacubes show no significant velocity differences between the emission clumps in Fig.~\ref{fig:recomb_map}; also the integrated line profiles are rather smooth and symmetrical. These characteristics suggest that the RTLs are not strongly masering, although weak amplification of the continuum may still be occurring (see below).

\section{Models and Discussion} \label{sec:model}

RTL $\alpha$ lines are formed by dipole transitions between $\Delta n=1$ energy levels that are strongly collisionally coupled with the next ionization state, so their emission strength depends on the abundance of that state. 

To examine whether the putative 232 GHz Rydberg transitions are consistent with
atmospheric conditions surrounding Betelgeuse we computed H30$\alpha$
and X30$\alpha$ line profiles using a semi-empirical thermodynamic atmospheric model. We have used an update of the model of \citet{2001ApJ...551.1073H}, to
include the revised distance and photospheric Rosseland angular diameter.
It uses the run of hydrogen ($n_{\rm H}$) and electron
($n_{\rm e}$) densities, and the
mean electron temperature (T$_{\rm e}$) with stellar radius derived from the VLA radio continuum visibilities from \cite{Lim1998}, a compilation of
published radio and mm-radio fluxes \citep{2001ApJ...551.1073H}, and {\it Hubble Space Telescope} ultraviolet (UV) fluxes
\citep{1994ApJ...428..329C}. This model represents the mean 1D dominant lukewarm chromosphere and not
the small filling factor of hot chromospheric plasma that gives rise to the rich UV spectrum. It also does not include pockets of cold molecular material. The outer reaches of the 2001 model were scaled to the
new distance as described in \citet[Sect. B1]{Harper2009}, and the inner region now
includes a spherical MARCS photospheric model with $T_{\mathrm{eff}}=3650$ and $\log{g_\star}=-0.5$
\citep{2008A&A...486..951G} with a turbulently extended chromospheric region.
Full details of the revised semi-empirical model will be published elsewhere, but the
radial dependence of $n_{\rm H}$ (which is predominantly neutral) and T$_{\rm e}$ are shown in the right panel of
Fig.~\ref{fig:radial_structure}.

\begin{figure}
\begin{center}
	\includegraphics[width=8.8cm]{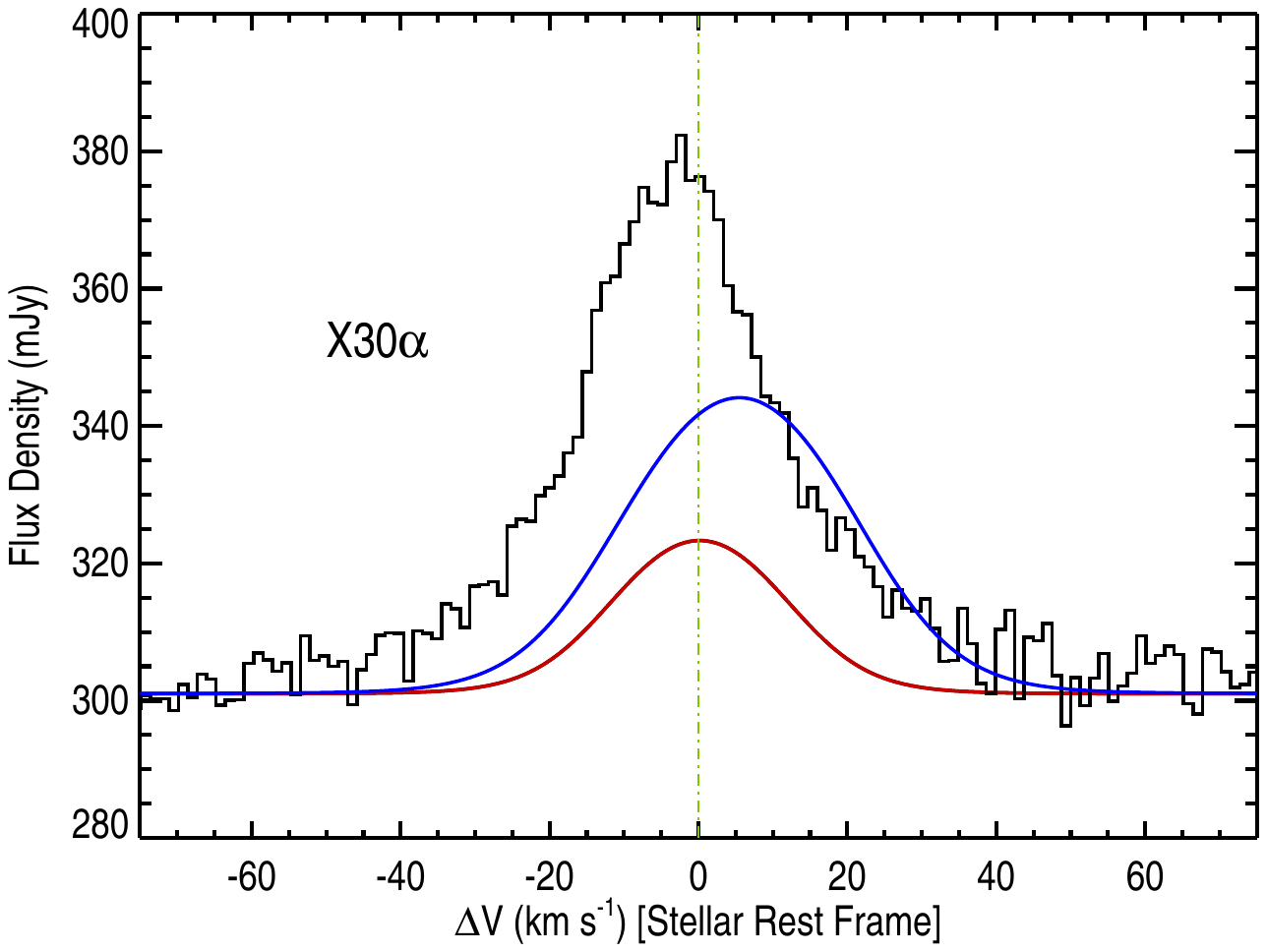}
    \includegraphics[width=8.8cm]{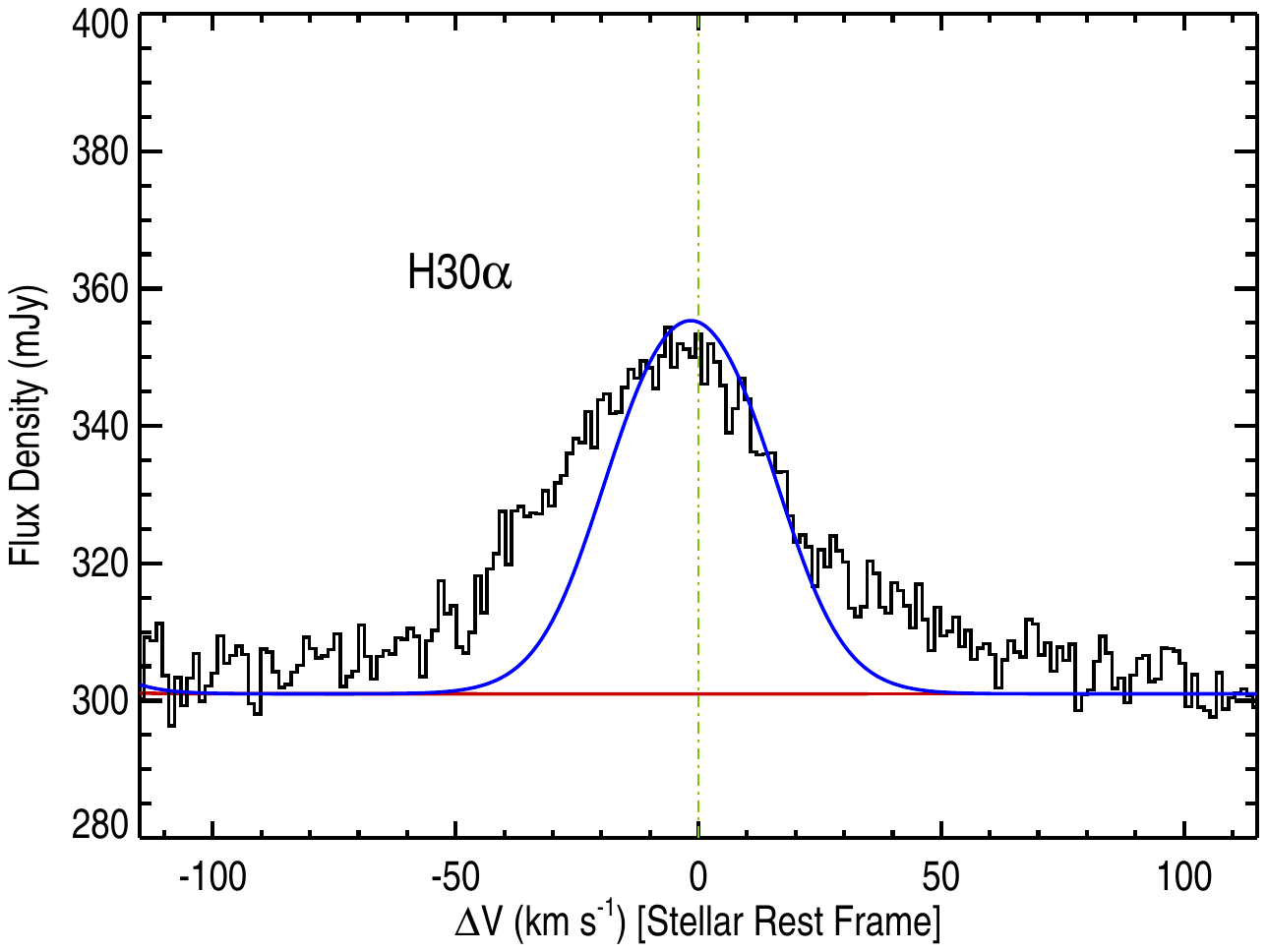}
    \caption{Simulations of the RTLs computed with and without enhanced C and H ionization.
     Left: X30$\alpha$ (232.023\,GHz). Mg, Si, and Fe  computed with the Saha-Boltzmann level populations producing $\sim 1/4$ of the observed flux (shown by the red line). In the second simulation (blue) the C is singly ionization and this improves the match of the total fluxes but shifts the combined profile redward by $\sim 5$\kmsns, further away from the observed peak. 
     Right: H30$\alpha$ (231.901\,GHz). In the first simulation no emission is predicted 
     (see text for details). We artificially increased the ionization fraction of H to $1\times 10^{-4}$ which now leads to H30$\alpha$ emission. Inclusion of non-LTE level populations will likely alter the predicted line fluxes. The observed lines, especially H30$\alpha$, show excess flux in the line wings.}
    \label{fig:X30sim}
    \end{center}
\end{figure}

The electron density in the RTL forming region in the lower chromosphere
is dominated by photoionized abundant low {\it First Ionization Potential} (FIP) metals,
namely Mg, Fe, and Si. The intense chromospheric UV 
radiation field that arises from the pervasive small filling factor of hot plasma can ionize
the low FIP elements and potentially some S and C, but not O whose ionization is tightly
coupled to neutral H by charge-exchange \cite[][and references therein]{Judge1986}. 
In the chromosphere, molecular hydrogen will have a negligible abundance. In the extended atmosphere to a first approximation if
Mg, Fe, Si are ionized then $n_{\rm e} \simeq 10^{-4} n_{\rm H}$, and if C and S are fully ionized then 
$n_{\rm e} \simeq 3\times 10^{-4} n_{\rm H}$.
To examine the RTLs observed both on and off the 232\,GHz stellar disk it was assumed the
upper and lower levels are hydrogenic and populated in their Boltzmann ratios.
The absolute level populations are derived from the Saha-Boltzmann equation from
the atmospheric hydrogen and electron densities. We adopt the
Fe and C abundances from \citet{2000ApJ...530..307C} with the isotopic ratio $^{12}{\rm C}/^{13}{\rm C}=6$ from
\cite{1984ApJ...284..223L}, and the Si abundance and isotopic
ratios given in \citet{2014A&A...561A..47O}. Solar abundances \citep{Asplund2021} were adopted for
Mg and S, and otherwise isotopic terrestrial isotropic ratios were used.

We computed the X30$\alpha$ cluster line profile with the following species
$^{12}{\rm C}$, $^{13}{\rm C}$, $^{23}{\rm Na}$, $^{24}{\rm Mg}$, $^{25}{\rm Mg}$, $^{26}{\rm Mg}$,
$^{28}{\rm Si}$, $^{29}{\rm Si}$, $^{30}{\rm Si}$, $^{32}{\rm S}$, and $^{56}{\rm Fe}$, in addition
to H30$\alpha$. The semi-empirical atmospheric model is based on continuum data and is agnostic to
atmospheric velocity fields, so we assume isotropic non-thermal motions ramping
from 11\kms at $1.3\,R_\star$ to 23\kms at $2.0\,R_\star$ (FWHM),
consistent with the observed widths of narrow (low opacity) far-ultraviolet emission lines
\citep{2018ApJ...869..157C}. The Doppler velocity in each line profile is the thermal motion
added in quadrature to the non-thermal motion. The rest frequencies were initially taken to be purely hydrogenic \citep[see][]{Towle1996}.
The continuum opacity sources included were
H~free-free, H$^{-}$ free-free and metal$^{-}$ free-free. The spherical radiative transfer problem,
a class of Local Thermodynamic Equilibrium (LTE) but with non-LTE ionization balances, was solved for the line and continuum. 

In the continuum the radial optical depth  $\tau_{232\>{\rm GHz}}$~=1 occurs near $1.3\>R_\star$, i.e., above the photosphere, and the X30$\alpha$ Rydberg emission occurs above this radius in the region
where the gas temperature and turbulence are increasing outwards. The 232\,GHz
stellar continuum flux density was computed to be $\simeq 300$\,mJy in good agreement with the observed value. The left panel of Fig.~\ref{fig:radial_structure} shows the computed disk size convolved with the ALMA beam (green dash line), also in good agreement with that observed (solid green), indicating that this is a reasonable model to compute the RTLs to establish their likelihood in Betelgeuse.

A comparison of the observed and model spectra is shown in Fig.~\ref{fig:X30sim}.
The left panel shows the computed X30$\alpha$ spectrum (red) where the electrons
come from Mg, Si, and Fe, and is plotted with respect to their abundance-weighted 
hydrogenic rest frequency (232.0232\,GHz) in the rest frame of the star. 
A second simulation (blue) with fully ionized C and S, still under-estimates the observed flux, but shifts the overall line profile red-ward by $\sim 5$\kmsns (away from the observed rest frequency) owing to the lower C atomic mass. 
In the right panel no hydrogen emission is produced in this model, as expected (see below); however the addition of a small hydrogen ionization fraction $(1\times 10^{-4})$ leads to H30$\alpha$ emission at its rest velocity, shown by the blue curve. 

We choose not to over-interpret the computed line fluxes because we have ignored non-LTE
level population effects that are likely to change opacities and source functions and
alter the distribution of the emission. For example, collisional and radiative cascades from higher n-levels and Mg II, and photoexcitation from the chromosphere and star need to be included to accurately predict the non-LTE level populations.
This might explain why the predicted X30$\alpha$ spatial distribution of the emission 
shown in Fig.~\ref{fig:radial_structure} is more extended than the observations.
We have explored schematic non-LTE simulations and find that masering is possible against the stellar disk and non-LTE effects are likely.
To investigate the uncertainty of the X30$\alpha$ rest frequencies we computed an l-state resolved hydrogenic Einstein A-value weighted frequency for $^{24}$Mg using the energy-level expressions in \cite{Chang1987}. This revealed a $\sim$1\kms blue-shifted frequency, and presumably similar uncertainties exist for the more complex Si and Fe atoms.

Both the X30$\alpha$ and H30$\alpha$ lines appear blue-shifted with respect to the model spectra, by 3 and 6\kms respectively. While the on-disk profile can easily be affected by inward and/or outward bulk motions, the annular emission which
samples the volume of atmosphere both in front of and behind the star should be less affected. The
ALMA emission shows no significant difference in the velocity centroid in the annular region. Although there is some uncertainty in the X30$\alpha$ central rest frequency (see above), the  H30$\alpha$ line has an essentially {\it exact} rest frequency and its off-disk spectrum is also blue-shifted. One possibility is some mild maser amplification of the stellar background continuum, along with net outward motion along the line-of-sight; this could also explain why the X30$\alpha$ line emission is more compact that predicted (see Fig.\ref{fig:radial_structure}). Further exploration of these line shifts and structure at higher spatial resolution and in different transitions would help elucidate these differences. 

The synthetic line profiles broadened by turbulence are similar to first order to those observed but lack the extended wings. Additional sources of line width have been discussed by \citet{Olofsson2021}; one possibility is collisional broadening - with electrons, ions or neutrals. The mean electron density measured from the $n_e$-sensitive C~II] 2325\AA ~emission multiplet in the hot chromospheric component is $\simeq 2\times 10^8\>{\rm cm}^{-3}$ \citep{Judge1998}, and \citet{Harper2006} show that the local density may reach $10^9\>{\rm cm}^{-3}$.  Assuming the approximation from \citet{Brocklehurst1972} (Eq.~4.8) is applicable to $30\alpha$ transitions, using T$_e$ = 3000K in the emitting region from the model in Fig.~\ref{fig:radial_structure}, the observed Lorentzian component of H30$\alpha$ in Table~\ref{tab:line_chars} would indicate an electron density of $\sim3\times 10^8\>{\rm cm}^{-3}$. Because of the strong dependence on the Rydberg transition, measurements of the widths of other lines would be of interest to confirm this mechanism.
Betelgeuse's measured surface longitudinal magnetic field is $\sim 1$\,G \citep{Auriere2010} which is insufficient to induce significant Zeeman broadening.

We do not expect hydrogen to have any noticeable ionization in the lukewarm chromosphere, but it
is very sensitive to small amounts of embedded hot plasma such as that which excites the
UV spectra. In a time-independent atmosphere H is thought to be ionized by a two-stage process: first the $n=2$ level is excited by electron collisions (which is very sensitive to the local temperature) and also by re-population by scattered H~Ly$\alpha$ photons in the massively opaque atmosphere, and
second, the n=2 level is then photoionized by the optically-thin photospheric Balmer continuum \citep{1984ApJ...284..238H}. In the simulation we arbitrarily added a small H ionization fraction 
in the model to create the emission shown in the blue spectrum. We note that gas 
that is periodically heated in shocks can be cool and over-ionized behind the 
shocks which might provide the additional ionization need to create H30$\alpha$ emission.  The clumpy nature of H30$\alpha$ in Fig.\ref{fig:recomb_map} compared with X30$\alpha$ suggests that its line-forming process is relatively unstable.
If Lorentzian wing broadening is a result of high electron densities then the H30$\alpha$ emission
(and some of the X30$\alpha$) must come from tiny pockets of hydrogen-ionized plasma, otherwise the 
entire radio flux spectrum and angular sizes would be discrepant by factors of many 
\citep{Harper2006}.

In summary, a plausible radial distribution of hydrogen and electron densities and
gas temperature in the extended atmosphere leads to reasonable X30$\alpha$ emission both on and
off the stellar disk confirming the identification of this emission. The presence of
H30$\alpha$ emission is readily explained by the presence of hot, perhaps
shocked, plasma within the atmosphere. Indeed small volumes
of hot plasma within the extended cooler material is the source of the
rich UV chromospheric emission spectrum.

\section{Conclusions}

We can confidently identify the observed emission lines around 232 GHz in the ALMA spectra of Betelgeuse as  H30$\alpha$ and X30$\alpha$. Synthetic spectra computed from a semi-empirical model atmosphere are in good agreement with the continuum and line observations considering the simple modeling assumptions adopted. Interesting velocity differences of a few \kms between the peaks of observed and modeled profiles remain unexplained. Line shapes show wings consistent with Voigt rather than Gaussian profiles, and electron densities within the hotter chromosphere
are consistent with collisional line broadening. In the future, Rydberg transition lines could prove to be valuable new diagnostics of the extended atmospheres of RSGs.

\begin{acknowledgments}

Support for GMH was provided by grant HST-GO-16256.001-A provided by Space Telescope Science Institute, which is operated by the Association of Universities for Research in Astronomy, Incorporated, under NASA contract NAS5- 26555.
This paper makes use of the following ALMA
data: ADS/JAO.ALMA\#2022.A.00026.S. ALMA is a partnership of ESO (representing its member states), NSF (USA) and NINS (Japan), together with NRC
(Canada), MOST and ASIAA (Taiwan), and KASI (Republic of Korea), in cooperation with the Republic of Chile. The Joint ALMA Observatory is operated by ESO, AUI/NRAO and NAOJ.
 PK acknowledges funding from the European Research Council (ERC) under the European Union's Horizon 2020 research and innovation program (project UniverScale, grant agreement 951549). LDM is supported by an award from the National Science Foundation (AST-2107681).

\end{acknowledgments}

%

\vspace{5mm}
\facilities{ALMA}






\bibliography{my_biblio}{}
\bibliographystyle{aasjournal}



\end{document}